# High-Performance Green and Blue Light-Emitting Diodes Enabled by CdZnSe/ZnS Core/Shell Colloidal Quantum Wells

*Yunke Zhu[1#], Xiuyuan Lu[4#], Jingjing Qiu[1#], Peng Bai[1,7]\*, An Hu[1], Yige Yao[1], Qinyun Liu[1], Yang Li[1], Wenjin Yu[1], Yaolong Li[1], Wangxiao Jin[4], Xitong Zhu[4], Yunzhou Deng[5], Zhetong Liu[6], Peng Gao[6], XiaoFei Zhao[7], Youqin Zhu[7], Li Zhou[7]\*, Yizheng Jin[4]\*, Yunan Gao[1,2,3]\**

1. State Key Laboratory for Mesoscopic Physics and Frontiers Science Center for Nano-optoelectronics, School of Physics, Peking University, Beijing 100871, China.
2. Collaborative Innovation Center of Extreme Optics, Shanxi University, Taiyuan 030006, China.
3. Peking University Yangtze Delta Institute of Optoelectronics, Nantong 226010, China
4. Key Laboratory of Excited-State Materials of Zhejiang Province, State Key Laboratory of Silicon and Advanced Semiconductor Materials, Department of Chemistry, Zhejiang University, Hangzhou 310027, China.
5. Cavendish Laboratory, University of Cambridge, CB3 0HE, Cambridge, UK.
6. Electron Microscopy Laboratory, School of Physics, Peking University, Beijing 100871, China.
7. BOE Technology Group Co., Ltd., Beijing 100176, China

\* Corresponding author
baipeng@pku.edu.cn
zhouli_cto@boe.com.cn
yizhengjin@zju.edu.cn;
gyn@pku.edu.cn;
\# These authors contributed equally to this work.

**Abstract**

The unique anisotropic properties of colloidal quantum wells (CQWs) make them highly promising as components in nanocrystal-based devices. However, the limited performance of green and blue light-emitting diodes (LEDs) based on CQWs has impeded their practical applications. In this study, we tailored alloy CdZnSe core CQWs with precise compositions via direct cation exchange (CE) from CdSe CQWs with specific size, shape, and crystal structure and utilized hot-injection shell (HIS) growth to synthesize CdZnSe/ZnS core/shell CQWs exhibiting exceptional optoelectronic characteristics. This approach enabled us to successfully fabricate green and blue LEDs manifesting superior performance compared to previously reported solution-processed CQW-LEDs. Our devices demonstrated a remarkable peak external quantum efficiency (20.4% for green and 10.6% for blue), accompanied by a maximum brightness 347,683 cd m$^{-2}$ for green and 38,063 cd m$^{-2}$ for blue. The high-performance represents a significant advancement for nanocrystal-based light-emitting diodes (Nc-LEDs) incorporating anisotropic nanocrystals. This work provides a comprehensive synthesis strategy for enhancing the efficiency of Nc-LEDs utilizing anisotropic nanocrystals.

**1. Introduction**

Colloidal quantum wells (CQWs), characterized by their anisotropic nature, are considered promising materials for display and laser applications due to their unique optoelectronic properties including ultranarrow emission width, controllable emission wavelength, giant oscillator strength and suppressed Auger recombination effect[1–16]. To enhance the performance of CQW-LEDs, various effective strategies including advanced chemical synthesis to improve the photoluminescence quantum yields (PLQYs) of CQWs, techniques of self-assembly for boosting out-coupling factor, and heterostructure tailoring of CQWs to achieve better carrier balance have been developed[2,3,17–21]. These efforts have resulted in the development of red CQW-LEDs with remarkably high external quantum efficiencies (EQEs) exceeding 25%[3], outperforming the majority of the colloidal quantum dot-based LEDs (CQD-LEDs). However, the progress of green and blue CQW-LEDs has been stagnant.

It is widely acknowledged that the incorporation of the inorganic shell with a wide-energy band gap is essential for enhancing the optical stability and PLQY of nanocrystals[12]. However, previous studies have demonstrated that the introduction of an inorganic shell protective layer to CQWs resulted in charge delocalization, leading to an excessive redshift in the emission wavelength to above 600 nm[20,22–24]. Despite previous efforts to suppress deep trap states and achieve high PLQYs at fixed emission wavelengths by employing core/alloyed-crown structures in green and blue Cd-based CQWs, the performance of green and blue devices remains unsatisfactory[5,7,9,15]. In general, these devices exhibit low electroluminescent (EL) efficiencies (EQE < 10%) and short operational lifetimes (the time when the luminance decays to 50% of the initial value of 100 cd m$^{-2}$, $T_{50}$@100 cd m$^{-2}$, < 1 h), as summarized in Table S2, thereby continuing to restrict the widespread application of devices.

Previous studies have successfully realized alloying wide bandgap materials, such as ZnSe or CdS (Figure S1), into Cadmium-based nanocrystals to effectively modulate their emission wavelengths[25–27]. This approach is expected to address the excessive redshift in the emission wavelength of core/shell CQWs. Nevertheless, direct synthesis of ternary CdZnSe CQWs has not yet been realized due to their unique reaction mechanism as anisotropic nanocrystals[28–31]. Recently, cation exchange (CE) reaction has been demonstrated as a successful strategy for CQWs to cover the entire visible gamut[22,32,33]. However, several unfavorable factors such as ligand-induced surface etching at high temperatures[22] and internal lattice defects due to copper residues from indirect CE reactions[32] can result in a diminished quantum efficiency of CQWs. These circumstances call for new strategies to develop CQWs to achieve high-performance green and blue CQW-LEDs.

In this study, we present a novel synthesis strategy for CdZnSe/ZnS core/shell CQWs covering the green and blue spectral range while exhibiting exceptional thermal stability and photoelectric properties. We successfully achieved the first direct cation exchange from Cd to Zn and controlled the Zn content in thin cadmium-based CQWs, while preserving their crystal structure and sheet-like morphology. By modulating the intrinsic bandgap of the core CQWs influenced by the compositions, we can address the excessive redshift in emission wavelength caused by subsequent shell growth[34]. The HIS approach can effectively passivate the alloy core CQWs, thereby simultaneously enhancing PLQY and suppressing deep trap channels. The emission peak can be finely adjusted within the range of 465 nm to 540 nm through controlled CE of CdZnSe alloy core CQWs and subsequent ZnS shell growth. By incorporating the CdZnSe/ZnS core/shell CQWs into the benchmark device structure, we have demonstrated superior performance of green and blue devices compared to previously reported solution-processed CQW-LEDs (Table S2).

## 2. Results and Discussion

**Synthesis of CdZnSe core CQWs with direct cation exchange reaction.** Direct cation exchange reactions, while maintaining the two-dimensional structure, were previously considered challenging for thin CdSe CQWs[32,33]. Ternary CdZnSe CQWs have only been achieved with a minimum thickness (>6 monolayers) or via indirect exchange with two steps facilitated by copper ions[32–34]. Here, we have developed a new strategy for subjecting 4.5 MLs (consisting of five Cd monolayers and four Se monolayers) CdSe CQWs to direct CE, successfully achieving the synthesis of well-preserved two-dimensional geometry in CdZnSe core CQWs (Figure 1a). Firstly, the exchange reactions are driven by disparate solubilities between the two cations by using ligands that preferentially coordinate to one of the metals. We utilized and modified the recipe reported by Li et al. for the sequential CE converting CdSe via CuSe to ZnSe nanocrystals[35], by introducing tributylphosphine (TBP), a widely utilized soft base[35–37], to solubilize the acidic $Cd^{2+}$ ions which are slightly softer than $Zn^{2+}$[38] according to the hard and soft acids and bases (HASB) theory[39]. The moderate extraction capacity of soft bases allows the reaction to proceed smoothly. Secondly, we used zinc halides, which exhibit a weaker binding capacity for oleylamine (OLA) compared to zinc oleate and zinc phosphonate[35], to efficiently release zinc ions, thereby displacing cadmium ions. OLA as a common ligand for CdSe nanocrystals helps to maintain the surface stability of the crystals and control the reaction rate to meet the condition of having reaction zone width smaller than the crystal size[36].

Figure 1b shows the solutions of CQWs after different CE reaction times (10 min to 60 min). As the CE reaction progresses, the composition of the nanocrystals undergoes effective changes, leading to a significant shift in the photoluminescence (PL) of the solution sample towards to the bluish color. The high-angle annular dark-field scanning transmission electron microscopy (HAADF-STEM) image (Figure 1c, Figure S2) confirms that the CQWs maintain a sheet-like morphology with a highly flat surface after the CE reaction. The excellent surface flatness allows the full width at half-maximum (FWHM) of the PL spectrum to remain below 20 nm (Figure 1d) with asymmetric broad emission tails attributed to trapping effects[40]. Trap state emissions are significantly suppressed after additional ligand passivation (Figure S3). The size of CQWs is reduced from 11.4 ± 0.9 nm to 10.2 ± 1.4 nm shown (Figure 1e) due to the etching[41] effect facilitated by OLA at a high temperature (~ 240°C). After the CE reaction, both the size distribution and shape of CQWs deteriorate, with their square shape transforming into irregular shape (Figure S2), indicating that the etching primarily occurs at the edges of the CQWs. Because of quantum confinement, an ensemble of CQWs exhibits varrying emission wavelengths[42], leading to inhomogeneous broadening (Figure 1d) caused by the inevitable size and shape distribution of CQWs[43].

The absorption and PL spectra (Figure 1d) exhibited continuous blue-shifting with increasing CE reaction times (10 min to 120 min), indicating the formation of CdZnSe ternary nanocrystals with a band gap wider than that of pure CdSe. Moreover, the emission wavelength of the CQWs can be tuned over a wide range from 512 nm to 455 nm. The absorption peaks of light-hole (lh) and heavy-hole (hh), which represent the properties of CQWs, remained clearly distinguishable, indicating that the lattice arrangement of the crystals remained undisturbed despite a slight increase. The nanocrystals were also characterized by powder X-ray diffraction (XRD, Figure 1f), revealing typical zinc blende (ZB) structure characteristics with distinct (111), (220), and (311) Bragg peaks observed following the CE reaction, accompanied by a noticeable shift towards higher angles.

The narrower intrinsic band gap of CdZnSe CQWs, in comparison to ZnSe CQWs with a zinc blende structure[44], implies an incomplete CE reaction and distinguishes it from previously observed phenomena of indirect exchange reactions[32,35,45]. The process of partial CE facilitates the conversion of preformed nanocrystals into either alloy nanocrystals or nanoheterostructures (NHCs) featuring core/shell or segmented architectures[38]. The uniform contrast of the nanocrystals precludes the formation of segmented NHCs. The previous study has reported that a critical temperature of 240 °C can be employed for the conversion of ZnSe/CdSe core/shell NHCs into $Zn_{1-x}Cd_xSe$ homogeneous alloy nanocrystals[46]. Alloyed CQWs can be inferred in our work since the CE reaction was performed at a temperature higher than the critical point. To verify the formation of alloy nanocrystals via Cd-to-Zn exchange, we used inductively coupled plasma-optical emission spectroscopy (ICP-OES) to analyze the compositional variations in the core CQWs. The Cd and Zn content in $Cd_xZn_{1-x}Se$ CQWs was quantified, revealing that approximately 53% of the Cd within the 4.5 MLs CQWs core had been substituted by Zn (Figure 2a, Table S1) after 10 min reaction. As the reaction progresses, the Zn content continues to increase, reaching ~ 82% after 180 min.

The exciton properties of the CdZnSe CQWs change significantly with time at different temperatures (Figure 2b, Figure S4). At high temperatures, Frenkel pairs, consisting of cation vacancy and self-interstitials, rapidly complete the exchange inside the ultrathin core CQWs [38], resulting in the formation of alloy nanocrystals. The formation of Frenkel defects is highly temperature-dependent, as they minimize the Gibbs free energy. The light-hole absorption peak undergoes only a slight blue shift at lower temperatures, while the ICP-OES results demonstrate that the CE reaction has been effectively carried out (Zn content > 30%, Table S1). We used the effective mass approximation[47] to calculate bandgap energy as a function of the Zn concentration (Figure 2c, Table S1), which aligns better with high-temperature reaction results and substantiates the formation of alloyed nanocrystals. In contrast, the CE reaction may occur only at the surface of the core CQWs at lower temperature due to insufficient formation of Frenkel pairs, consistent with previous work[34].

Trioctylphosphine (TOP), another tertiary phosphine, also exhibits the capability of serving as a soft base ligand in CE reactions (Figure S5). Despite having similar hardness[38], TOP has a larger site resistance than TBP, allowing it to efficiently solvate the harder $Zn^{2+}$ in the nanocrystals. In addition to the soft base phosphine, excess OLA used to solubilize the cationic precursor binds readily to the nanocrystal surface, acts as a passivator and hindering the CE reaction (Figure S6). Consequently, our work signifies the initial implementation of the CE reaction for core CQWs, providing a potential new approach for producing low-cadmium and cadmium-free CQWs in future applications.

**Synthesis of CdZnSe/ZnS core/shell CQWs.** Capping nanocrystals with an inorganic protective shell has been recognized as essential for enhancing their optoelectronic properties, such as PLQY and stability, thereby meeting the requirements for LEDs and lasers[12]. To achieve efficient green and blue CQW emitters, we synthesized $Cd_{1-x}Zn_xSe$/ZnS core/shell heterostructure based on the CdZnSe alloyed cores by hot-injection shell (HIS) approach[48] (Figure 3a). The excitons in CQWs can be strongly limited by the large band gap of ZnS, preventing non-radiative recombination by passivating surface defects of the CQWs[17].

We set the reaction temperature to 305 °C to break the carbon-sulfur bonds in 1-octanethiol, releasing sulfur. The resulting green (G) and blue (B) CdZnSe/ZnS CQWs exhibited a distinct nanoplate morphology, with uniform sizes of ~ (12.2 ± 1.4) nm and (12.6 ± 1.4) nm, respectively (Figure 3a, below). The enlarged absorption cross sections demonstrate the formation of the core/shell CQWs, while showing PL peaks at 524 nm (G) and 474 nm (B), as shown in Figure 3b. After shell growth, the emission of the CQWs is moderately red-shifted (Figure 3c) due to the delocalization of the carrier wave function[49] and achieved an extremely narrow FWHM (~ 24 nm (G) and 18 nm (B)), thus meeting the requirements for display and lighting applications. The as-synthesized CQWs (Figure S7) exhibit a distinct "edge-up" orientation (thin but bold appearance) or a "face-down" configuration (roughly rectangular uniform contrast). We used back focal plane imaging[50] to examine the transitions dipole moment (TDM) distribution of the CQW films spin-coated on top of the hole transfer layer (poly(9-vinylcarbazole)). We found that the TDM distribution of green (~ 77%) and blue (~ 76%) CQWs films was larger than the isotropic CQDs (~ 67%), indicating the preparation of anisotropic nanocrystals (Figure S8).

The type-I heterostructure substantially extends the average PL decay lifetime of the CQWs from less than 5 ns to 16.2 ns and 13.4 ns (Figure 3d, Figure S9), with the suppression of the short-lived channel further confirming the growth of an inorganic protective layer with passivating effects[21,40]. Consistent with the change trends of the average lifetime, the PLQY of CdZnSe/ZnS CQWs efficiently increased compared to CdZnSe core CQWs (Figure 3e, Figure S9), which is crucial for preparing high-performance devices. The wide bandgap inorganic shell and stable thiol ligands confer solution stability, maintaining 90% of the initial PLQY after five ethanol purifications. Our synthesis strategy effectively improves the optoelectronic properties of nanocrystals.

It is important to point out that the influence of ligands is also crucial for the growth of these heterostructure nanocrystals. During the synthesis, we varied the quantities of ligands like Oleic acid (OA) and OLA in order to investigate the product characteristics. The carboxylic acid ligand OA not only stabilizes the crystal surface and facilitates nanocrystal growth but also reacts in situ with added zinc acetate to generate the desired cationic source, zinc oleate. Varying the amount of oleic acid affects the optimal growth in the vertical direction of the CQWs, leading to changes in exciton behavior (Figure S10), consistent with previous reports[48]. In addition, introducing OLA modifies the nanocrystal morphology, resulting in the presence of CQDs in the final product. This phenomenon is attributed to subsequent digestive ripening occurring after the decomposition of CQWs through OLA etching (Figure S11).

**Device fabrication and performance.** The high quality green and blue CdZnSe/ZnS core/shell CQWs provided a solid material foundation for developing CQW-LEDs with outstanding performance. We fabricated CQW-LEDs with a benchmark structure based on CQWs as EML with excellent photoelectric properties. The devices (Figure 4a) consisted of multiple layers in the following order, indium tin oxide (ITO)-coated glass substrates, poly(ethylenedioxythiophene): polystyrene sulphonate (PEDOT:PSS, ≈ 30 nm), poly(9-vinylcarbazole) (PVK, ≈ 30 nm), CdZnSe/ZnS core/shell CQWs (≈ 10 nm), ZnMgO (≈ 60 nm), and silver electrodes (≈ 80 nm). We take advantage of the deep high-occupied-molecular-orbit (HOMO) energy level of PVK, the common hole-transport material for CQD-LEDs, to realize efficient hole injection into the CQWs layer. The green and blue CQW-LEDs achieved record peak EQEs of 20.6% and 10.4%, respectively. The histogram of 20 devices indicates an average peak EQE of 19.6% and 9.4% with a low relative standard deviation of 0.8% and 0.6% (Figure 4c). With the high-performance EML, the green and blue devices in this batch achieved maximum luminescence of 347,683 cd m$^{-2}$ and 38,063 cd m$^{-2}$, respectively (Figure 4d). As shown in Figure 4e, the electroluminescence spectrum of the device hardly changed until the voltage rises to 9 V, with Commission Internationale de l'Eclairage (CIE) coordinates showing minimal variation with increasing luminance (Figure S14). At a display-related luminance of 1000 cd m$^{-2}$, the CIE coordinates are (0.195, 0.741) and (0.108, 0.156). Combined with our previous work[3], the color gamut of CQW-LEDs covers 151% of sRGB, and exceeds 82% of Rec.2020 standard. The green and blue CQW-LEDs exhibit a Lambertian emission profile (Figure S15), which is important for display applications.

The operational lifetime of previous CQW-LEDs has typically been poor. For example, CQW-LEDs with a similar structure were reported with a lifetime of only 249 h at an initial luminance of 100 cd m$^{-2}$[13]. Studies on the lifetime of blue and green devices are even less reported. In this context, we achieved devices based on CdZnSe/ZnS core/shell CQWs with long operational lifetimes (Figure 5, Figure S16). The $T_{50}$ lifetime at an initial luminance of 51,990 cd m$^{-2}$ of a green device is determined to be 28 min, corresponding to a $T_{50}$@100 cd m$^{-2}$ of ~14,000 h. The blue CQW-LED realized a $T_{50}$ lifetime of 4,670 cd m$^{-2}$ of ~ 17 min, corresponding to a $T_{50}$@100 cd m$^{-2}$ of ~ 195 h. This is almost the first report on the stability of green and blue CQW-LEDs, though both lifetimes are still much shorter compared to those of CQD-LEDs[12]. The overall performance of our green and blue devices surpasses that of previously reported CQW-LEDs, as evidenced by a comprehensive comparison of key metrics including efficiency, luminance, and lifetime (Table S2).

## 3. Conclusion

We have successfully synthesized CdZnSe alloy core CQWs for the first time via the phosphine ligand-catalyzed direct cation exchange reaction. Based on these unique cores, we synthesized CdZnSe/ZnS core/shell CQWs via HIS growth and achieved electroluminescent green and blue CQW-LEDs with record efficiency and high brightness. Each performance parameter of the green and blue devices shows orders of magnitude improvement over the previous best results. Although the performance is still slightly lacking compared to the best-performing Nc-LEDs, this work fills the wide gap of the previous poor performance of blue and green CQW-LEDs and makes possible the application of various anisotropic nanocrystal-based devices. Moreover, it provides a pioneering guide for the utilization of anisotropic nanocrystals in fields of anisotropic nanocrystals and nanocrystal-based display.

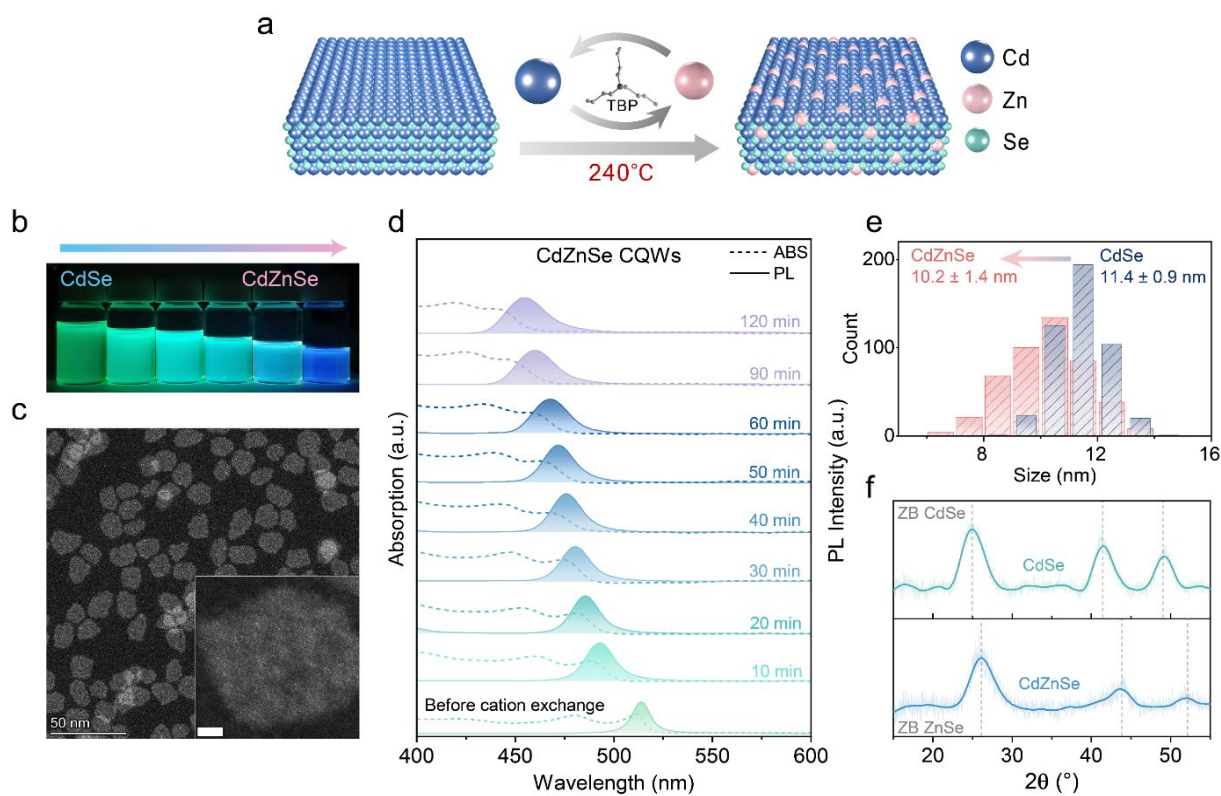

**Figure 1.** a) A schematic of the synthesis pathway from CdSe to CdZnSe CQWs. (b) Solutions of CQWs after different CE reaction times (10 min to 60 min) under 365 nm UV light. (c) HAADF-STEM images of CQWs after CE reaction with 40 min. The inset shows the high-resolution image of the CQW (scale bar, 5 nm). (d) Normalized absorption and PL spectra of CdZnSe CQWs with respect to the CE reaction time. (e) Size distribution histograms of CdSe and CdZnSe (40min) CQWs. f) XRD profiles of CdSe and CdZnSe (40 min) CQWs.

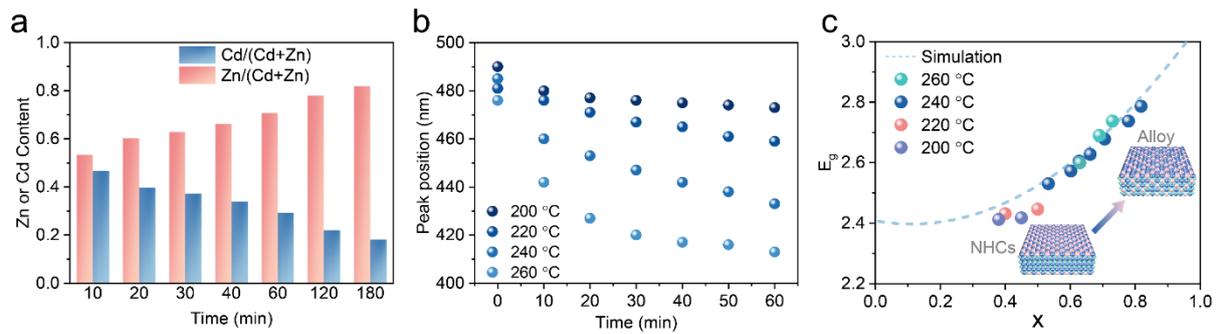

**Figure 2.** a) Relationship between cation ratio within the $Cd_{1-x}Zn_xSe$ core CQWs and reaction time. b) Absorption wavelength of light-hole during the CE reaction depending on the reaction time. c) Relationship between optical gap cation ratio within the $Cd_{1-x}Zn_xSe$ core CQWs after CE reaction at different temperatures and Zn content. The dashed line indicates composition dependence optical gap of alloyed $Cd_{1-x}Zn_xSe$ CQWs. Insets, corresponding to the Schematic illustrations of evolution from core/shell NHCs to alloy nanocrystals.

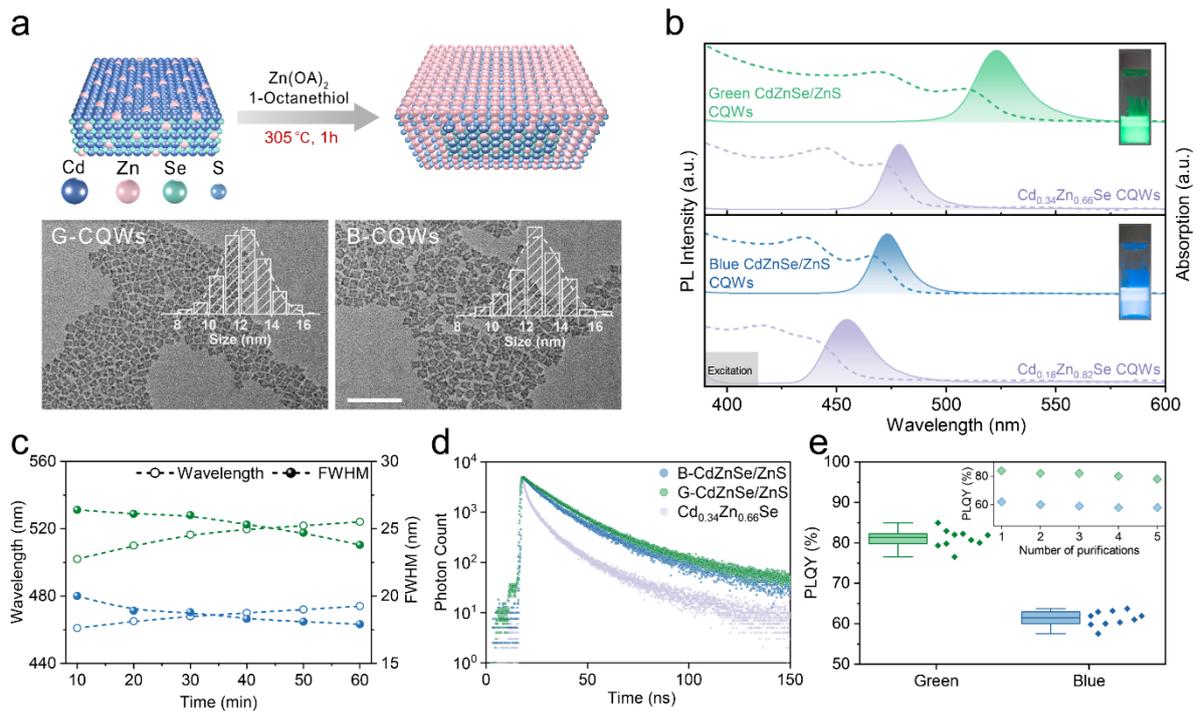

**Figure 3.** a) Schematic illustrations of the synthesis of CdZnSe/ZnS core/shell CQWs (upper panel). TEM images of green and blue CQWs (lower panel), (scale bar, 100 nm). Insets, corresponding size distribution histogram. b) Absorption and photoluminescence spectra of CQWs solution. Insets show photographs of the sample of core/shell CQWs under UV light. b) TEM images of green (above) and blue (below) CdZnSe/ZnS core/shell CQWs. c) Evolution of wavelength and FWHM of the PL peak during the HIS growth. d) Ensemble PL decays of CQWs. e) The PLQY of green and blue CQWs solutions. Insets, corresponding to the purification test of the CQWs.

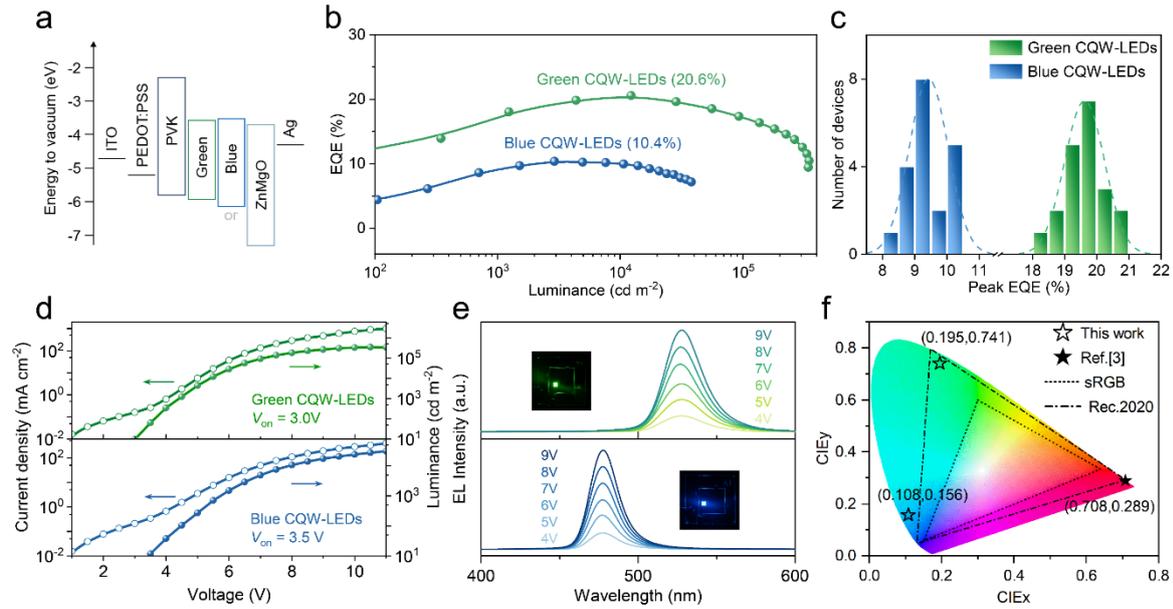

**Figure 4.** a) Schematic energy level diagram of multilayer CQW-LED structure. b) EQE-Luminance relationship of blue and green device. c) Histogram of peak EQEs from 20 devices. d) Current density-luminance-voltage characteristics of blue and green devices. e) EL spectra of CQW-LEDs under different biases. The inset shows the operating CQW-LEDs. f) The CIE 1931 coordinates of green and blue CQW-LEDs. The coordinates of the best-reported red CQW-LEDs[3].

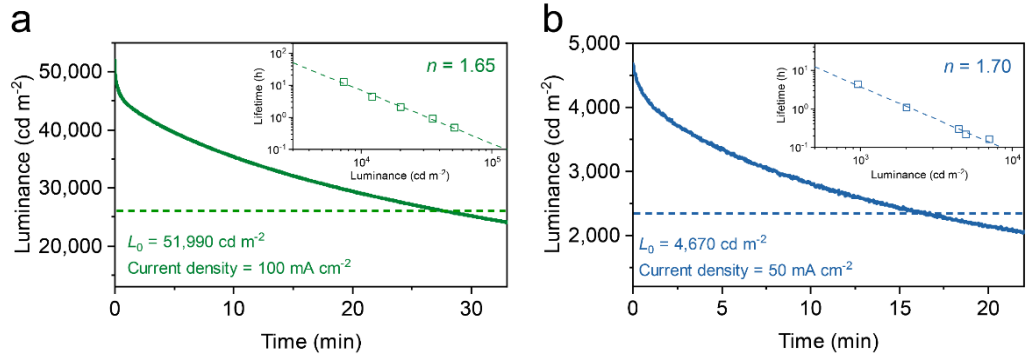

**Figure 5.** Luminance as a function of operational time for green (**a**) and blue (**b**) CQW-LED. Device lifetime is evaluated by measuring luminance over time at constant current density. The lifetimes ($T_{50}$) at various initial luminance ($L_0$) values are shown in the insets. The acceleration factors ($n$) are fitted according to the empirical relationship of $(L_0)^n T_{95}$ = constant.